# Strain Engineering for Phosphorene: The Potential Application as a Photocatalyst

Baisheng Sa[1], Yan-Ling Li[2,*], Jingshan Qi[2], Rajeev Ahuja[3] and Zhimei Sun[4]

[1] College of Materials Science and Engineering, Fuzhou University, Fuzhou 350100, P. R. China
[2] School of Physics and Electronic Engineering, Jiangsu Normal University, 221116, Xuzhou, People's Republic of China;
[3] Condensed Matter Theory Group, Department of Physics and Astronomy, Uppsala University, Box 516, 75120 Uppsala, Sweden and Applied Materials Physics, Department of Materials and Engineering, Royal Institute of Technology, 10044 Stockholm, Sweden
[4] School of Materials Science and Engineering, and Center for Integrated Computational Materials Engineering, International Research Institute for Multidisciplinary Science, Beihang University, 100191 Beijing, P. R. China

**Abstract**

Phosphorene has been attracted intense interest due to its unexpected high carrier mobility and distinguished anisotropic optoelectronic and electronic properties. In this work, we unraveled strain engineered phosphorene as a photocatalyst in the application of water splitting hydrogen production based on density functional theory calculations. Lattice dynamic calculations demonstrated the stability for such kind of artificial materials under different strains. The phosphorene lattice is unstable under compression strains and could be crashed. Whereas, phosphorene lattice shows very good stability under tensile strains. Further guarantee of the stability of phosphorene in liquid water is studied by *ab initio* molecular dynamics simulations. Tunable band gap from 1.54 eV at ambient condition to 1.82 eV under tensile strains for phosphorene is evaluated using parameter-free hybrid functional calculations. Appropriate band gaps and band edge alignments at certain pH demonstrate the potential application of phosphorene as a sufficiently efficient photocatalyst for visible light water splitting. We found that the strained phosphorene exhibits significantly improved photocatalytic properties under visible-light irradiation by calculating optical absorption spectra. Negative splitting energy of absorbed $H_2O$ indicates the water splitting on phosphorene is energy favorable both without and with strains.

*To whom all correspondence should be addressed: ylli@jsnu.edu.cn.

**INTRODUCTION**

The discovery of graphene with the two dimension (2D) hexagonal honeycomb carbon gives rise to the research hotspot of 2D materials in recent years[1-2]. Graphene-based and graphene-related 2D materials have gained increasing interest as the visible-light photocatalyst in the water splitting hydrogen production process[3-4]. Apart from graphene, many binary 2D materials, for instance, single layer boron-nitride[5], monolayer $MoS_2$[6] as well as related 2D heterostructures[7-8] have been evaluated to be potential visible-light photocatalysts with suitable electronic structures. The energy levels of the conduction and valence bands and the size of the band gap are vital points in semiconductor photocatalysts: the conduction band minimum (CBM) should be located more negative than the redox potential of $H^+/H_2$ (0 V vs NHE), and the valence band maximum (VBM) should be occupied more positive than the redox potential of $O_2/H_2O$ (1.23 V vs NHE). As a result, the band gap for a photocatalyst should be larger than 1.23 eV with suitable band edge alignment[9].

Recently, a new mono-elementary 2D material, named layered black phosphorus or phosphorene has been fabricated[10-11] and attracted global interest due to its unexpected anisotropic optoelectronic and electronic properties[12-18]. For example, phosphorene's high carrier mobility have been theoretically predicted[19] and experimentally measured[20-21]. Negative Poisson's ratio[22] and enormous mechanical flexibility[23] have been found in phosphorene as well, indicating the potential application of phosphorene at extreme conditions. Phosphorene was predicted to be a nearly direct gap semiconductor with 0.7~2 eV evaluated band gap[10, 19, 24-27]. Combined with the 1.45 eV experimentally photoluminescence (PL) spectra peak[10], phosphorene shows the potential possibility as a photocatalysts.

Phosphorene can theoretically hold up to 30% critical strain[23], signaling the possibility of tuning its physical and chemical properties by strain engineering[28-31]. Very recently, Rodin *et al*[32]. found the transition from semiconductor to metal in phosphorene by applied inter-planar strains. Fei *et al*[33]. discovered the preferred conducting direction in phosphorene can be rotated by appropriate intra-planar biaxial or uniaxial strains. Peng *et al*.[34] explored the strain induced direct to indirect band gap transition mechanism in phosphorene. Fei *et al*[35]. raveled the strain induced lattice vibrational modes and Raman scattering shifts in phosphorene. Furthermore, Elahi *et al*[36]. unfolded the emergence of Dirac-shaped band structure dispersion in phosphorene under certain strain conditions. These studies show that electronic structure of phosphorene is very sensitive to the applied strain. Hence, combing with reasonable band gap of phosphorene, we believe the studies of the photocatalysis-related properties of phosphorene are of great interest and importance under strain engineering.

Herein, we systematically studied the crystal and electronic structures of phosphorene for the applications as a photocatalyst under uniaxial intra-planar strains based on the density functional theory calculations. We unraveled the band gap and band edge alignment modifications under no more than 10 % compression and tensile strains. Moreover, the lattice dynamically stability of phosphorene under certain strains were estimated as well. Furthermore, we assumed that this remarkable 2D material will find its applications as the visible-light photocatalyst in the water splitting hydrogen production.

## COMPUTATIONAL DETAILS

We studied phosphorene based on density functional theory calculations. The Vienna *ab initio* simulation package[37] (VASP) in conjunction with projector augmented wave (PAW) pseudopotentials within the generalized gradient approximations[38] (GGA) of Perdew-Burke-Ernzerhof[39] (PBE) was used. The valence electron configuration for P was $3s^23p^3$. The geometry convergence was achieved with the cut-off energy of 500 eV. K-points of 8×10×3 were automatically generated with the Γ symmetry. The relaxation convergence for ions and electrons were $1\times10^{-5}$ and $1\times10^{-6}$ eV, respectively. The PHONOPY[40] code was applied to get the phonon frequencies through the supercell approach with a 3×6×1 supercell and 2×2×1 K-points. Since the phosphorene layers are bonded by the van der Waals forces in bulk black phosphorous[41], the van der Waals corrected optB88 function[42] was also introduced for all the calculations comparison. We found that the van der Waals interactions in the phosphorene are less insignificance than the bulk black phosphorous, most of the optB88 results agree with the PBE very well. Moreover, we introduced the parameter-free Heyd-Scuseria-Ernzerhof (HSE06) hybrid functional[43] to precisely evaluate the band gap of phosphorene. We performed $G_0W_0$ and $GW_0$ calculations to check the band gap of phosphorene as well.

## RESULTS AND DISCUSSIONS

**The Crystal Structure of Phosphorene under Strain.**

The optimized lattice parameters, P-P bond length and bond angle for phosphorene are listed in Table 1. The lattice parameters $a$ = 4.626 Å, $b$ = 3.298 Å using PBE and $a$ = 4.506 Å, $b$ = 3.303 Å using optB88, respectively. We have also analyzed the intra-planar and inter-planar P-P bond length and bond angle. According to Fig.1, the optimized intra-planar bond length and bond angle for phosphorene is $R_1$ = 2.220 Å, $\theta_1$ = 104.18 ° using PBE, and $R_1$ = 2.226 Å, $\theta_1$ = 103.04 ° using optB88, respectively. The inter-planar P-P bond length and bond angle for phosphorene is $R_2$ = 2.259 Å, $\theta_2$

= 95.92 ° using PBE, and $R_2$ = 2.260 Å, $\theta_2$ = 95.80 ° using optB88, respectively. Our results agrees well with the previous studies very well[19,33]. It is clear that PBE and optB88 describe the inter-planar parameters and inter-planar bond length in a very similar way. At the same time, it exhibits about 1 ° differences between the PBE and optB88 inter-planar bond angle. As a result, optB88 reproduce smaller lattice parameter *a* value than PBE. Anyway, the errors are less than 2.6 %, which indicates that the van der Waals interactions dose not play an important role in the monolayer phosphorene. Hence we present the PBE results in the following part.

In order to study the intra-planar strain engineering for phosphorene, we have applied up to 10% compression and tensile uniaxial strains along the *a* axis (the armchair direction) and *b* axis (the zigzag direction). Fig. 2 illustrates the P-P bond length and bond angle as a function of the applied strain. The blue and red marks show the P-P bond length and bond angle under strain along *a* and *b* axis, respectively. It is understandable that the compression and tensile strain influence the phosphorene lattice and crystal structure in the opposite direction. More interestingly, we found that phosphorene shows opposite tendency under strain along different axis. Take the tensile strain as an example: the intra-planar P-P bond length $R_1$ shows tiny response to the strain along *a* axis, but is remarkable influenced by the strain along *b* axis. The inter-planar P-P bond length $R_2$ will be increased with the increasing of the tensile strain along *a* axis, which will be decreased according to the tensile strain along *b* axis. As the tensile strain along *a* axis increases, the intra-planar bond angle $\theta_1$ significantly increases and the inter-planar angle $\theta_2$ slightly decreases. As the tensile strain along *b* axis increases, the intra-planar bond angle $\theta_1$ slightly decreases and the inter-planar angle $\theta_2$ significantly increases. And for the compression strain, vise verse.

**The Lattice Dynamical Stability of Phosphorene under Strain.**

In order to explore the lattice dynamical stability, we have calculated the phonon dispersion curves. Fig. 3 illustrated our calculated phonon dispersion curves for phosphorene at ambient condition, which agrees well with the former studies[35-36]. We have further studied the sound velocities by fitting the slopes of the acoustic dispersion curves around the Γ point. As can be seen from Table 1, the speed of longitudinal and transverse sound along the Γ to X direction are $v_{LA}^{\Gamma-X}$ = 3.96 km/s and $v_{TA}^{\Gamma-X}$ = 3.61 km/s. The speed of longitudinal and transverse sound along the Γ to Y direction are $v_{LA}^{\Gamma-Y}$ = 7.99 km/s and $v_{TA}^{\Gamma-Y}$ = 3.95 km/s. Our results excellently agrees with the maximum velocity of sound $v_{max}^{\Gamma-X}$ = 3.8 km/s along Γ to X and $v_{max}^{\Gamma-Y}$ = 7.8 km/s along Γ to Y by Zhu *et al*[25]. As seen from Fig.3, no negative or imaginary frequency was found, suggesting that phosphorene shows good lattice dynamical stability at ambient condition.

Figs. 4 and 5 plot the phonon dispersion curves for phosphorene under different uniaxial strains. For both the situations, compression strains will introduce imaginary frequency dispersion into phosphorene, which leads to lattice instability. Interestingly, the imaginary frequency dispersions behave different under different uniaxial strains. For the strain along *a* axis, the phosphorene lattice remains stable under small compression strains (see Fig. 4 (d)). As the compression strain increased (see Fig. 4 (c) and (b)), the imaginary part of the acoustic phonon mode appears and grows along the Γ to X direction as well as the Γ to Y direction. These imaginary acoustic mode is very similar to the imaginary dispersion curves found in $ThH_2$[44]、$Bi_2Te_3$[45] and $Ge_2Sb_2Te_5$[46], which can be easily enhanced to positive by the temperature effects or temperature induced electron phonon interactions. It is worth noting that another imaginary optical mode appears under more than 9 % applied compression strain along *a* axis, as is shown in Fig. 4 (a). Such a negative dispersion mode may lead to the crash of the phosphorene lattice. For the strain along *b* axis, small compression strains firstly lead to negative acoustic phonon mode along the Γ to Y direction (see Fig. 5 (d)). As the strain increased, the imaginary part grows and one acoustic phonon mode along the Γ to X direction will be negative as well (shown in Fig. 5 (c) and (b)). The phenomenon is corresponding to those cases under strain along *a* axis. However, the strain along *b* axis will not lead to any imaginary optical mode. On the contrary, one double degenerated acoustic phonon mode is negative around the M point in Fig. 5 (a) under 10 % applied compression strain along *b* axis, which may lead to the crash of the phosphorene lattice as well. In a word, phosphorene under compression strains is not suitable for the photocatalysis applications because of its lattice instability. Meanwhile, according to Fig. 4 (e), (f) and Fig. 5 (e), (f), phosphorene shows very good lattice stability under all tensile strains, which agrees well with the enormous strain limit of phosphorene[23]. Hence the strain free or stretched phosphorene meets the stability requirement for the photocatalysis applications.

**The Electronic Structure of Phosphorene under Strain and Photocatalysis Properties Analysis.**

Monolayer phosphorene has been theoretical predicted to be a nearly direct band gap semiconductor[24, 32]. However, since the energy difference between the actual VBM and the VBM at the Γ point is less than 10 meV[24]. We can briefly consider phosphorene as a direct gap semiconductor. Table 2 lists the calculated band gap at the Γ point using different methods. Some previously computational results[10, 19, 24-27] as well as the estimated value from the experimental absorption peak[10] are listed for comparison. It is well known that PBE calculations normally underestimate the electronic band gap, and the hybrid function with the mixing of the Hartree-Fock and DFT exchange terms is thought to be a practical solution to solve the band gap problem[47-48]. Herein, we have obtained 1.54 eV band gap for phosphorene using HSE06 hybrid functional based on PBE method. The corresponding result using HSE06 hybrid functional based on optB88 method is 1.49 eV. Both the values agree well with Qiao *et al.*'s 1.51 eV band gap using HSE06 at optB88 method[19], Tran *et al.*'s 1.4 eV band gap using BSE method[24], Rudenko *et al.*'s 1.6 eV band gap using GW method[27] and 1.45 eV experimentally PL spectra peak[10].

Since PBE and optB88 method represent very similar features and tendencies, we will only present the results based on PBE in the following. Fig. 6 (a) illustrated the partial density of states (DOS) for phosphorene. It is clearly that the covalent bond in phosphorene is mainly contributed by the *p-p* bonding states at the valence band and partially *s-p* antibonding states at the conduction band. It is noted that HSE06 hybrid functional enlarges the band gap of phosphorene without any other significant changing of the electronic structures. Hence we present detailed band structure analysis based on the PBE results. The band structure dispersion curve and the projected band structure using PBE have been plotted in Fig. 6 (b). The irreducible representations (IR) for the band around the Fermi level at the Γ point were marked as well. It is clear that both the VBM $\Gamma_2^+$ state and the CBM $\Gamma_3^-$ state are occupied by the P $p_z$ electrons. Another notable state at the Γ point is the $\Gamma_2^+$ state occupied by the P *s* electrons at the conduction band, which occupies the second lowest conduction band and could be switched to CBM under certain strain[33].

We have found that HSE06 and PBE methods present the same band gap changing tendency of phosphorene under strain, where the HSE06 gap is about 0.6 eV larger than the PBE gap. Fig. 6 (c) illustrated the calculated real band gap and direct band gap at the Γ point for phosphorene under strains using HSE06 functional. Our HSE06 results reproduced the direct to indirect band gap transition behaviors, which have been raveled by Peng et al.[34] using PBE functional. The compression strains shows no benefit to the photocatalysis properties by further reducing the phosphorene band gap far from the visible light range. At the same time, considering that the compression strains will introduce instability to the phosphorene lattice from our previously analysis, we focused on the band gap variation under tensile strains only. For the strain along *a* axis, the band gap of phosphorene firstly increases with the increasing of tensile strain. The size of the band gap peaks at 1.79 eV under 7 % tensile strain. The band gap slightly decreases when the strain keeps increasing. At the same time, the VBM transfers from the Γ point to the Y point, and thus phosphorene will be transformed into an indirect band gap semiconductor, which agrees well with Peng *et al.*[34]'s analysis. Such Γ to Y indirect band gap charge transfer may lead to low efficiency energy conversion by solar energy with additional lattice dynamic behaviors, which is not good for the photocatalysis applications. For the strain along *b* axis, the band gap of phosphorene increases first and then decreases as well. The maximum band gap is 1.82 eV under 5 % tensile strain. Unlike the strain along *a* axis case, large tensile strain along *b* axis lower the band gap without transfer phosphorene to be an indirect band gap semiconductor. The band gap drops sharply with the increasing tensile strain along *b* axis, since the CBM is occupied by the P *s* electrons dominated $\Gamma_2^+$ state instead of P $p_z$ electrons dominated $\Gamma_3^-$ state. Nevertheless, the band gap values are larger than required 1.23 eV minimum band gap for the photocatalysis reactions, showing the potential application of phosphorene as a visible light photocatalyst.

Fig. 6 (d) plots the band edge alignments of phosphorene with respect to the Normal Hydrogen Electrode (NHE). The standard water reduction and oxidation potential levels were marked for reference. At ambient condition, CBM of phosphorene locates more negative than the redox potential of $H^+/H_2$ (0 V vs NHE), but the VBM of phosphorene does not be occupied more positive than the redox potential of $O_2/H_2O$ (1.23 V vs NHE). Although tensile strains along both the directions slightly reduce the VBM positions, the VBM of phosphorene is more negative than the water oxidation potential in the whole strain range. Hence, phosphorene is not suitable for water splitting under vacuum conditions. Nevertheless, in a photocatalysis water splitting device, the redox potential for water depends on the pH value of the solutions. the standard oxidation potential $O_2/H_2O$ in a solution is[49]:

$$E^{ox}_{O_2/H_2O} = -5.67eV + pH \times 0.059eV, \qquad (1)$$

which could shift the water's oxidation potential upward in Fig 6 (d). According to Eq. (1), by changing pH, we can tune the band edge alignment of phosphorene suitable to the redox potential of $H^+/H_2$ and $O_2/H_2O$. Moreover, we can adjust the oxidizing power (defined as the difference between VBM and the oxygen reduction potential) and the reducing power (defined as the difference between CBM and the hydrogen reduction potential) as well, which protects the chemical balance between the reduction and oxidation reaction. For instance, in a pH = 8.0 solution, the oxidation potential as well as the reduction potential will shift upward 0.472 V in Fig 6 (d). On the basis of the above results, Fig. 7 illustrates the energy alignment of phosphorene in pH = 8.0 solutions. As is seen, phosphorene shows a favorable band position for water splitting both at ambient condition and under tensile strains in pH = 8.0 solutions. For all the cases we have shown, the VBM locates more positive than the $H^+/H_2$, and the CBM is more negative than the $O_2/H_2O$ potential, the feature of which meets the requirement of a photocatalyst for water splitting. Moreover, phosphorene has 1.76 eV band gap under 7 % tensile strain along the *a* axis and 1.82 eV band gap under 5 % tensile strain along the *b* axis. Such sufficient band gaps locate in the visible light wavelength range could harvest the visible light in a very high-efficiency way. Hence the strain engineering improves the photocatalysis properties by adjusting the band gap.

Fig. 8 plots the calculated optical spectra as a function of light wavelength for the cases shown in Fig. 7. It is seen that phosphorene for the cases show substantial adsorption both in the visible light and UV light range. Moreover, one can see that the adsorption of phosphorene under tensile strains in the visible-light region (dashed line part) is more noticeable than the strain free phosphorene. This is because sufficient 1.79 eV and 1.82 eV band gaps under strains of phosphorene absorb the visible light more efficiency, which is located in the visible light wavelength range.

Since the photocatalytic hydrogen production process starts from the splitting of absorbed water. We firstly evaluated the driving force of the splitting of water molecular on phosphorene $E_{\text{splitting}}^{M}$ in Table 3 by the following equations:

$$E_{\text{splitting}}^{\text{H}^+} = E_{\text{H}^+:\text{phosphorene}} + E_{\text{OH}^-} - E_{\text{H}_2\text{O:phosphorene}}, \tag{2}$$

$$E_{\text{splitting}}^{\text{OH}^-} = E_{\text{OH}^-:\text{phosphorene}} + E_{\text{H}_3\text{O}^+} - (E_{\text{H}_2\text{O:phosphorene}} + E_{\text{H}_2\text{O}}), \tag{3}$$

$$E_{\text{splitting}}^{\text{H}} = E_{\text{H:phosphorene}} + E_{\text{OH}} - E_{\text{H}_2\text{O:phosphorene}}, \text{ and} \tag{4}$$

$$E_{\text{splitting}}^{\text{OH}} = E_{\text{OH:phosphorene}} + E_{\text{H}} - E_{\text{H}_2\text{O:phosphorene}}, \tag{5}$$

where $E_{M:\text{phosphorene}}$ shows the total energy of phosphorene with water splitting product $M$ ($M$ = H$^+$, OH$^-$, H and OH). The corresponding splitting energy for the ionic H$^+$, OH$^-$ and charge free H, OH are -4.183, -0.704, -2.225 and -2.962 eV/H$_2$O, respectively. Negative results protect that the water splitting process on phosphorene is energy favorable. Moreover, as can be seen in Table 3, appropriate tensile strains further increase the driving force of the splitting of water with more negative $E_{\text{splitting}}^{M}$.

To further explore the water splitting process on phosphorene, we studied the charge transfer from water splitting product $M$ to phosphorene by analysis the charge density differences $\rho_{\text{diff}}$, which is obtained by:

$$\rho_{\text{diff}} = \rho_{M:\text{phosphorene}} - (\rho_{M} + \rho_{\text{phosphorene}}) \tag{6}$$

where $\rho_{M:\text{phosphorene}}$, $\rho_{M}$ and $\rho_{\text{phosphorene}}$ corresponding to the self-consistent charge density of relaxed $M$ absorbed phosphorene, the water splitting product $M$ and phosphorene, respectively. Fig. 9 illustrates $\rho_{\text{diff}}$ for different $M$ on phosphorene. The violet isosurfaces present the charge depletion $\rho_{\text{diff}} < 0$, the cyan isosurfaces show the charge accumulation $\rho_{\text{diff}} > 0$. According to Figs. 9 (a) and (b), the angle between the P-H bond and the phosphorene plane is bigger than the P-H$^{+}$ bond and the phosphorene plane. Otherwise, similar charge transfer features have been found in H and H$^{+}$ on phosphorene. As seen in Figs. 9 (c) and (d), the absorbed OH$^{-}$ ion accumulate more charges than charge free OH. Interestingly, the absorption of H or H$^{+}$ on one side of phosphorene will introduce the charge accumulation on the other side of phosphorene. Conversely, the absorption of OH or OH$^{-}$ on one side of phosphorene will lead to the charge depletion on the other side. The opposite charge transfer features for H/H$^{+}$ and OH/OH$^{-}$ absorptions indicate that the water oxidation and reduction process will take place separately on different side of monolayer phosphorene.

**The Stability of Phosphorene in Liquid Water.**

Last but not least, we evaluated the stability of phosphorene in aqueous environment by means of *ab initio* molecular dynamics (AIMD) simulations. Herein, phosphorene monolayer was placed into liquid water with fixed density of 1 g/cm$^{3}$ and annealed at the temperature of 300 K for 5 ps. It is clearly seen from the structure snapshot after annealing in Fig. 10 (a), the strain free phosphorene layer remains stable in liquid water, where the P atoms vibrate around their equilibrium positions under the combined effect of water and temperature field. The admirable stability of tensile strained phosphorene is proved by identically AIMD calculations. To give a further evaluation, we show in Fig. 10 (b) the normalized pair correlation functions $g(r)$ for phosphorene and water after annealing at 300 K. The P-P pair correlation function in Fig. 10 (b) verifies the phosphorene layer is well crystallized. The $H_2O$ pair correlation function shows the liquid nature of water, agrees well with XRD measurements[50].

**CONCLUSION**

In summary, we have systematically studied phosphorene for the applications as a photocatalyst under strains based on the density functional theory calculations. The anisotropy structural evolutions under different strains were first investigated. We found that phosphorene is lattice dynamical instable under compression strains, but it shows very good stability under tensile strains. The stability of phosphorene in liquid water has been further confirmed by AIMD calculations. Appropriate band gap and band edge alignment at certain pH ignites the potential application of phosphorene as a visible light photocatalyst. The strain engineering improves the photocatalysis properties by adjusting the band gap of phosphorene into the visible light wavelength range, which will facilitate the absorption of the visible light and thus increase the efficiency of the photocatalytic water splitting. The water splitting process on phosphorene is energy favorable, and the water oxidation and reduction process will take place separately on different side of monolayer phosphorene.

**ACKNOWLEDGMENTS**

This work was supported by the National Natural Science Foundation of China (Grant No. 11347007 and No. 61274005), Qing Lan Project, the Priority Academic Program Development of Jiangsu Higher Education Institutions (PAPD) and the National Natural Science Foundation for Distinguished Young Scientists of China (Grant No. 51225205). R. A. thanks the Swedish Research Council (VR) and Swedish Energy Agency for financial support.

Tabel 1. The calculated lattice parameters, P-P bond length, bond angle and velocity of sound for phosphorene using PBE and optB88 method.

| Parameters | PBE | optB88 |
|---|---|---|
| a (Å) | 4.626 | 4.506 |
| c (Å) | 3.298 | 3.303 |
| $R_1$ (Å) | 2.22 | 2.226 |
| $R_2$ (Å) | 2.259 | 2.26 |
| $\theta_1$ (°) | 104.18 | 103.04 |
| $\theta_1$ (°) | 95.92 | 95.80 |
| $v_{LA}^{\Gamma-X}$ (km/s) | 3.96 | 4.12 |
| $v_{TA}^{\Gamma-X}$ (km/s) | 3.61 | 3.87 |
| $v_{LA}^{\Gamma-Y}$ (km/s) | 7.99 | 7.98 |
| $v_{TA}^{\Gamma-Y}$ (km/s) | 3.95 | 4.25 |

Table 2. The calculated band gap for phosphorene.

| Method | Band Gap |
|---|---|
| PBE | 0.92 |
| HSE06@PBE | 1.54 |
| optB88 | 0.76 |
| HSE06@optB88 | 1.49 |
| $G_0W_0$@PBE | 2.08 |
| $GW_0$@PBE | 2.29 |
| GGA[25] | 0.7 |
| PBE[26] | 0.9 |
| HSE06@optB88[19] | 1.51 |
| modified HSE06@PBE[10] | 1.0 |
| $G_0W_0$[24] | 2.0 |
| BSE[24] | 1.2~1.4 |
| GW[27] | 1.6 |
| expt. absorb. peak[10] | 1.45 |

Table 3. The calculated driving force of the splitting of water molecular on phosphorene.

| $E^{M}_{splitting}$ (eV) | strain free | $a/a_0$=1.07 | $b/b_0$=1.05 |
|---|---|---|---|
| $E^{H^+}_{splitting}$ | -4.183 | -4.204 | -4.278 |
| $E^{OH^-}_{splitting}$ | -0.704 | -1.042 | -0.918 |
| $E^{H}_{splitting}$ | -2.225 | -2.470 | -2.254 |
| $E^{OH}_{splitting}$ | -2.458 | -2.612 | -2.575 |

Fig. 1

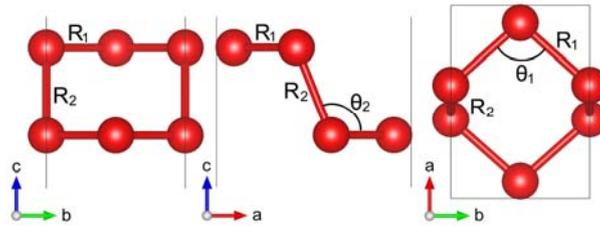

Fig. 1. The crystal structure of phosphorene. $R_1$ and $R_2$ indicate the intra-planar and inter-planar P-P bond length, $\theta_1$ and $\theta_2$ show the intra-planar and inter-planar bond angle.

Fig. 2

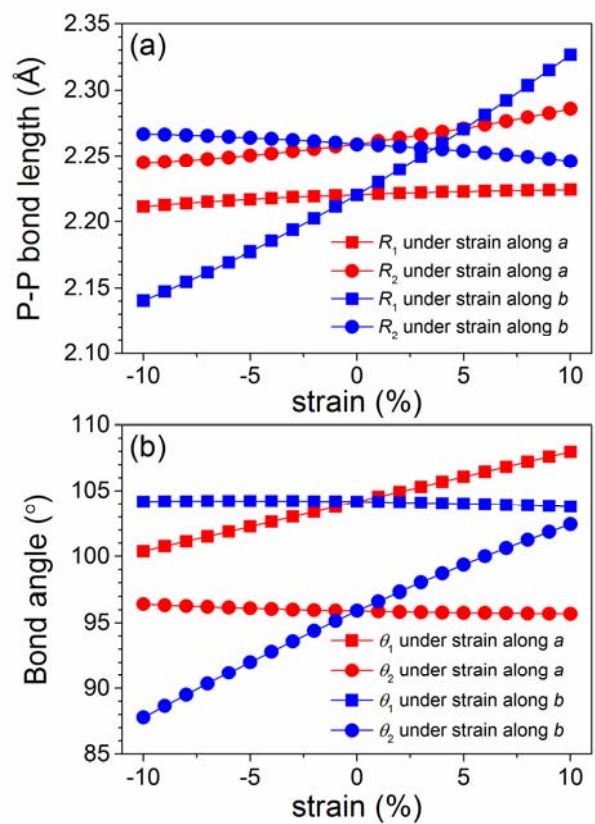

Fig. 2. (a) The P-P bond length and (b) bond angle as a function of the uniaxial strain.

Fig. 3

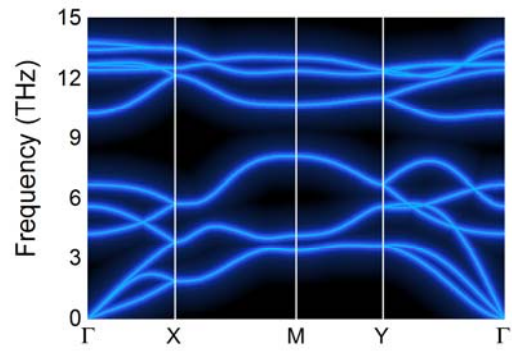

Fig. 3. The phonon dispersion curves of phosphorene.

Fig. 4

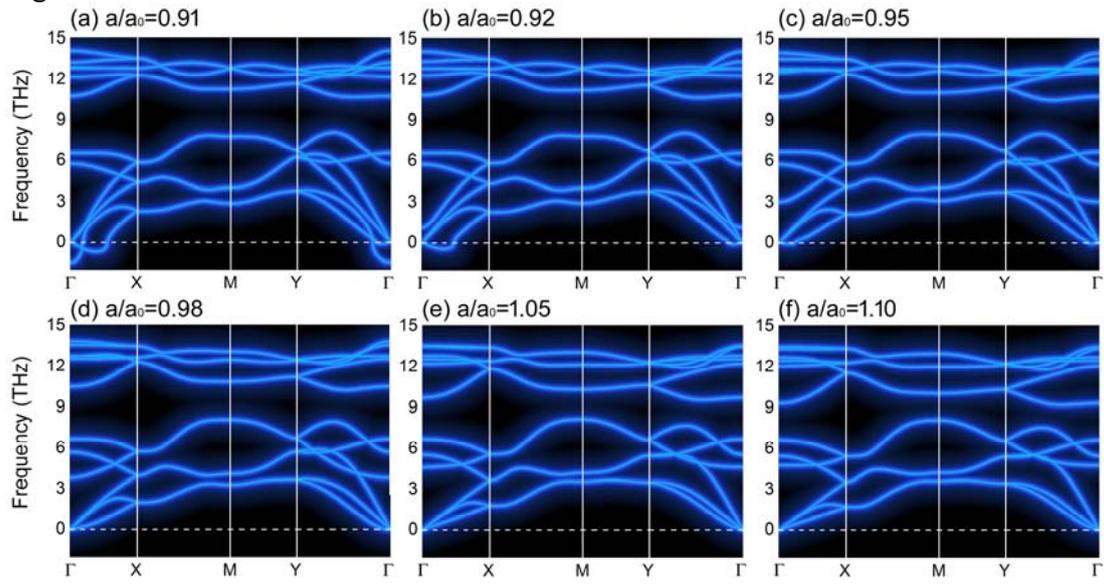

Fig. 4. The phonon dispersion curves of phosphorene under (a) 9 %, (b) 8 %, (c) 5 %, (d) 2 % compression strain and (e) 5 % and 10 % tensile strain along *a* axis.

Fig. 5

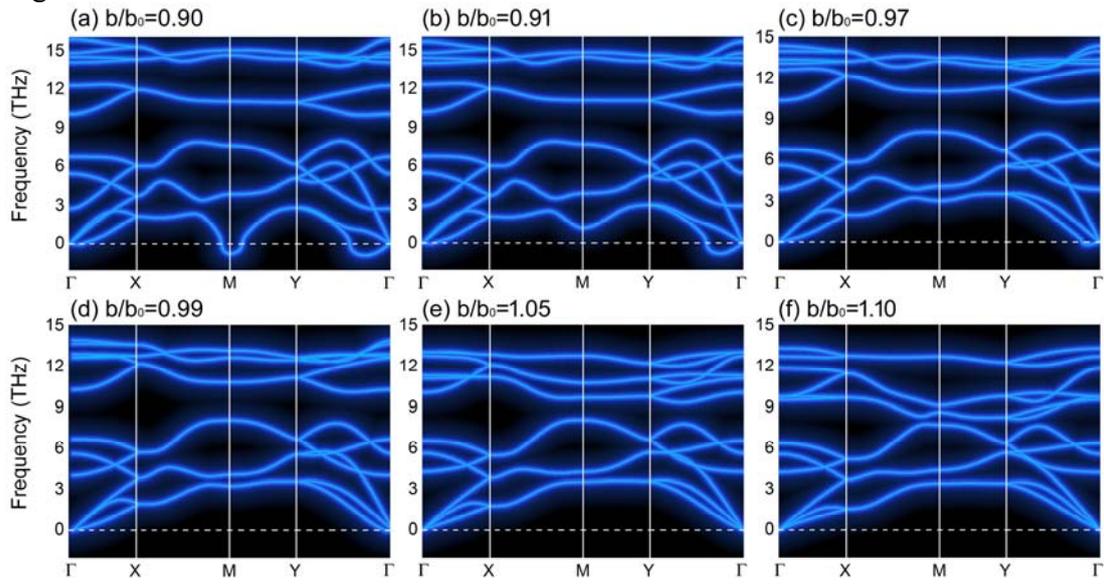

Fig. 5. The phonon dispersion curves of phosphorene under (a) 10 %, (b) 9 %, (c) 3 %, (d) 1 % compression strain and (e) 5 % and 10 % tensile strain along *b* axis.

Fig. 6

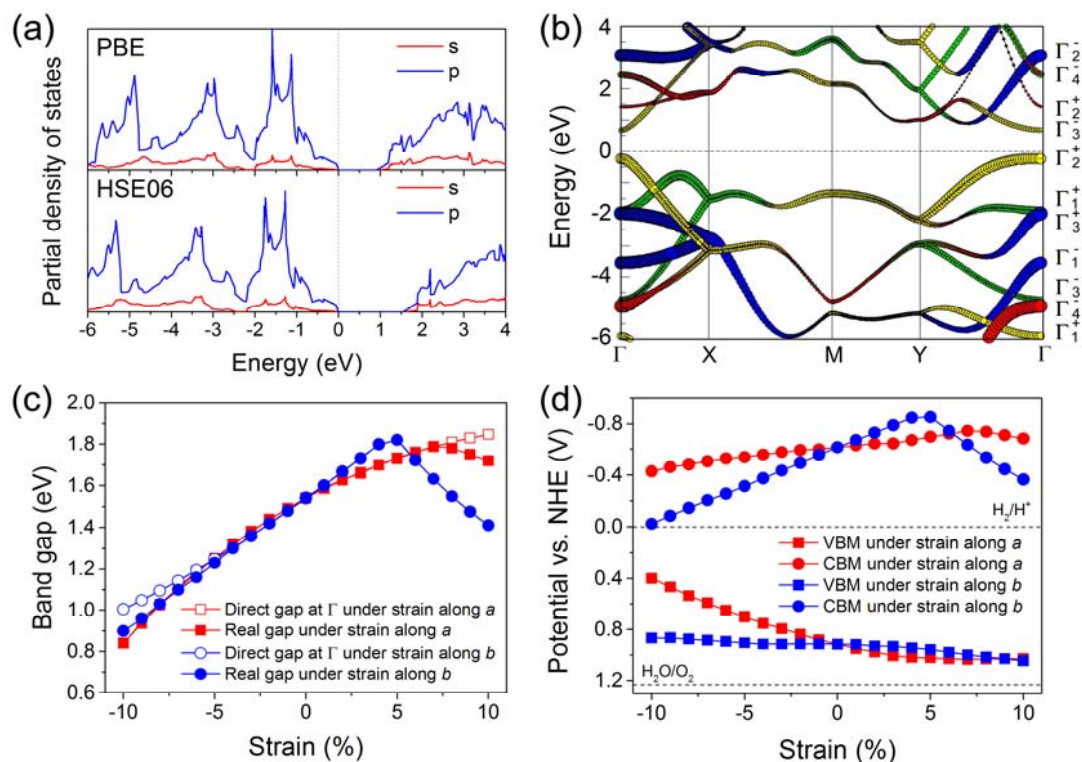

Fig. 6. (a) The partial density of states of phosphorene using PBE and HSE06. (b) The band structure plot of phosphorene using PBE. The Fermi energy is set to 0 eV. The symbols present the irreducible representations (IR) for the band around the Fermi level at the Γ point. The size of the red, green, blue and yellow circles illustrates the projected weight of P $s$, $p_x$, $p_y$ and $p_z$ electrons. (c) The band gap of phosphorene as a function of the uniaxial strain using HSE06 functional. (d) The evolution of the VBM and CBM of phosphorene as a function of the uniaxial strain using HSE06 functional. The dashed lines are standard water redox potentials. The reference potential is the vacuum level.

Fig. 7

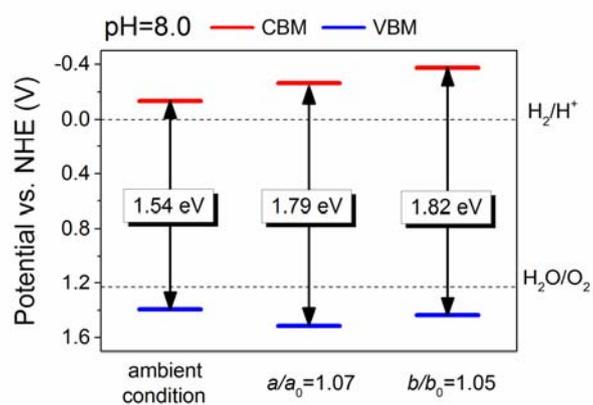

Fig. 7. The energy alignment of phosphorene at ambient condition, under 5 % tensile strain along *a* axis and under 6 % tensile strain along *b* axis when pH = 8.0. The dashed lines are water redox potentials in pH = 8.0 solutions.

Fig. 8

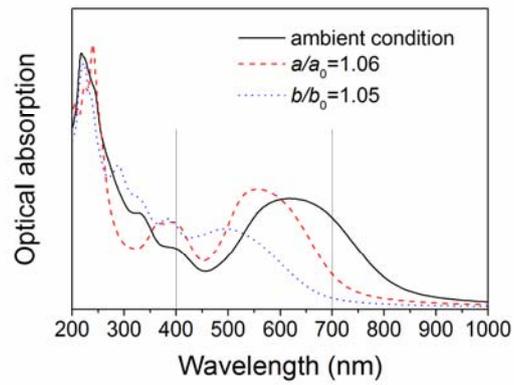

Fig. 8. The optical spectra for phosphorene at ambient condition, under 5 % tensile strain along *a* axis and under 6 % tensile strain along *b* axis, where the solid vertical lines show the wavelength range of visible light.

Fig. 9.

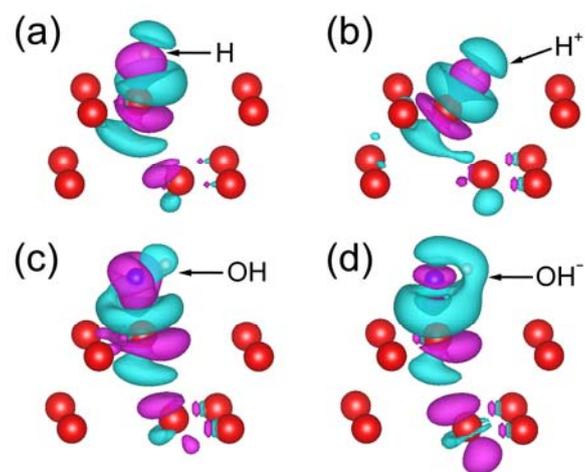

Fig. 9. The plot of charge density difference (violet: charge depletion, cyan: charge accumulation) for (a) H, (b) $H^+$, (c) OH and (d) $OH^-$ absorbed on top of phosphorene.

Fig. 10.

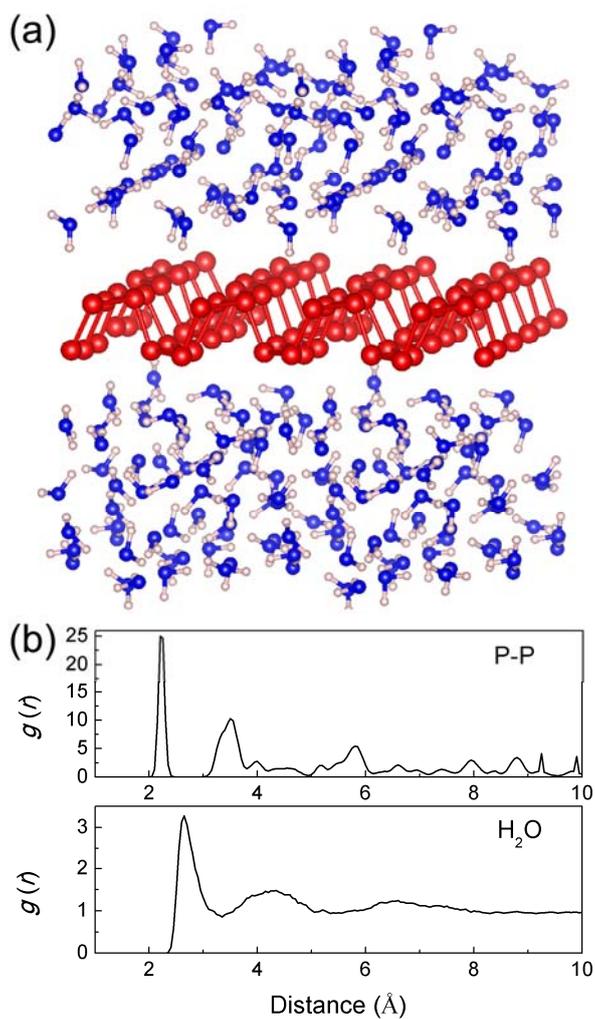

Fig. 10. (a) The structure evolution of strain free phosphorene in liquid water after annealing at 300 K. The red, blue and pink balls present the P, O and H atom, respectively. (b) The normalized pair correlation functions $g(r)$ after annealing at 300 K.

Table of Contents Graphic

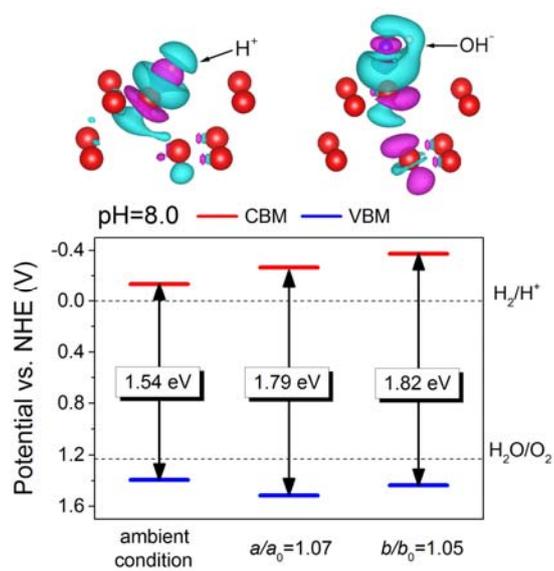